# SAM Molecular Stacking with Heterogeneous Orientation for High-Performance Perovskite Photovoltaics

Lei Huang,[1] Kai-Li Wang,[1,*] Zhang Chen,[1] Zhen-Huang Su,[2] Saidjafar Murodzoda,[4] Xin Chen,[1] Jing Chen,[1] Chun-Hao Chen,[1] Yu Xia,[1] Yu-Tong Yang,[1] Jia-Cheng Li,[1] Dilshod Nematov,[4] Ilhan Yavuz,[3,*] and Zhao-Kui Wang[1,5,*]

[1] State Key Laboratory of Bioinspired Interfacial Materials Science, Institute of Functional Nano & Soft Materials (FUNSOM), Jiangsu Key Laboratory of Advanced Negative Carbon Technologies, Soochow University, Suzhou 215123, China

[2] Shanghai Synchrotron Radiation Facility, Shanghai Advanced Research Institute, Shanghai Institute of Applied Physics Chinese Academy of Sciences, Shanghai 201204, China

[3] Department of Physics, Marmara University, Ziverbey, Kadikoy, Istanbul 34722, Turkiye

[4] S.U. Umarov Physical-Technical Institute of the National Academy of Sciences of Tajikistan, Dushanbe 734063, Tajikistan

[5] Lead Contact

*Correspondence Author: Z.K.W. (email: zkwang@suda.edu.cn), I.Y. (email: ilhan.yavuz@marmara.edu.tr), K.L.W. (email: klwang@suda.edu.cn)

# ToC

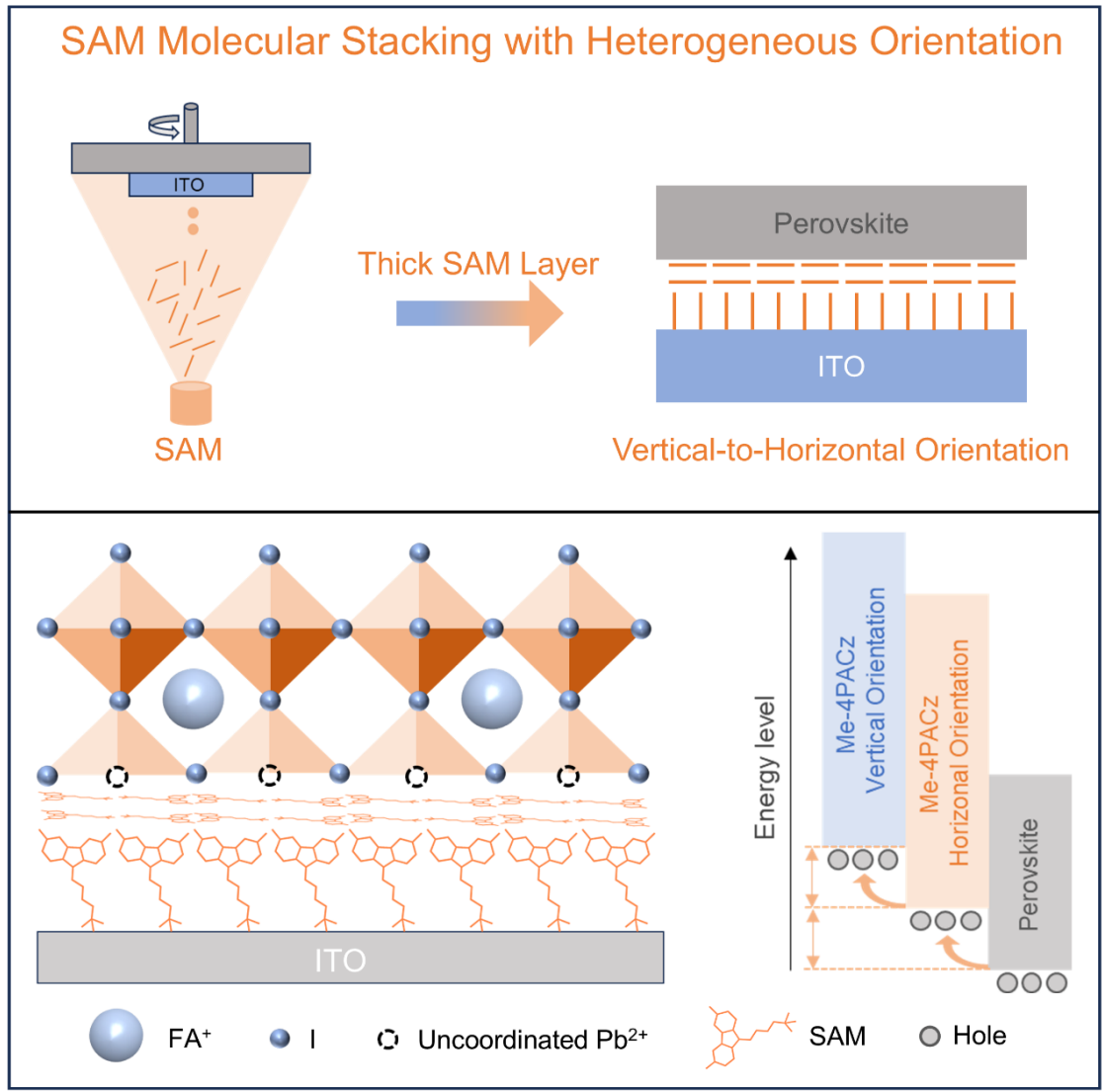

This study developed a molecular packing strategy with heterogeneous orientation for self-assembled monolayer (SAM) materials in perovskite photovoltaics. The evaporated SAM thick films spontaneously form a vertical-to-horizontal gradient in molecular orientation with excellent surface coverage. This heterogeneity gives rise to a gradient energy barrier that optimally facilitates hole transport. Capitalizing on the inherent benefits of thermal evaporation, the newly proposed strategy is readily applicable for fabricating large-area uniform SAM films, thereby overcoming a key challenge in achieving homogeneous SAM deposition for scalable production with traditional solution-based techniques.

# SUMMARY

The inability to achieve uniform hole transport with solution-processed self-assembled monolayers (SAM) constitutes a fundamental bottleneck for scaling perovskite photovoltaics. Herein, we demonstrate that thermal-evaporated SAM (eSAM) overcome this by enabling precise thickness control. Crucially, a thickened eSAM spontaneously forms a vertical-to-horizontal gradient in molecular orientation, which creates a descending energy barrier that directionally facilitates hole transport. This tailored interface also ensures excellent surface coverage and directs the growth of high-quality perovskite films. Consequently, the resultant photovoltaic devices set new benchmarks, delivering impressive power conversion efficiencies (PCEs) of 21.46% (small-area, 0.108 cm$^2$) and 19.38% (large-area module, 15.52 cm$^2$) for fully vacuum-evaporated devices, while also setting an impressive PCE of 23.67% for eSAM-based devices with solution-processed perovskites. The unencapsulated devices also present excellent operational stability by retaining 91.9% initial efficiency after 1200 hours continuous MPPT testing. The new strategy effectively addresses the critical challenge of scalable SAM deposition, positioning eSAM as a key enabler for the industrial advancement of perovskite photovoltaics.

**Keywords:**
Self-assembled monolayer (SAM); Heterogeneous molecular orientation; Hole transport; Perovskite photovoltaics.

# INTRODUCTION

Perovskite solar cells (PSCs) represent a breakthrough in photovoltaics, offering the favorable combination of high efficiency, low production cost, and adjustable optoelectronic features for next-generation applications.[1-6] In recent years, the incorporation of self-assembled monolayer (SAM) materials for hole transport has significantly advanced the performance of PSCs.[7-10] However, challenges such as molecular aggregation and non-uniform film coverage have hindered their effectiveness, particularly in large-area fabrication.[11-15] Thus, developing scalable SAM deposition methods is critical for the industrial application of perovskite photovoltaics.

Thermal-evaporated SAM (eSAM) can leverage the technique inherent strengths, including uniform deposition and precise thickness control, which makes it suitable for large-scale, homogeneous thin-film production.[16-19] In fact, the deposition thickness is particularly critical to the final device performance.[20-24] Currently, the underlying transport mechanism associated with the thickness of eSAM remains poorly understood. A thorough investigation into how film thickness influences hole transport dynamics is essential to fully realize the potential of eSAM in high-efficiency solar cells. In our work, the prominent SAM material Me-4PACz was utilized to elucidate the hole transport behavior governed by its thickness. We found that thick eSAM layer (Thick eSAM) spontaneously forms a heterogeneous molecular orientation transitioning from vertical to horizontal, thereby creating a graded energy barrier at the buried interface. This structure facilitates more efficient charge transport compared to the single energy barrier formed in thin eSAM layer (Thin eSAM). In addition, the Thick eSAM provides excellent molecular coverage on both the ITO and perovskite surface, mitigating the poor molecular alignment typical of Thin eSAM. Furthermore, due to the horizontal orientation of the upper molecules for Thick eSAM, the exposed phosphonic acid groups can passivate undercoordinated $Pb^{2+}$ defects at the perovskite buried interface to suppress non-radiative recombination while promoting hole transport efficiency.

Based on the study of the thickness mechanism under eSAM, we applied Thick eSAM to both fully vacuum-deposited and solution hybrid device structures. Remarkably, fully vacuum-deposited PSCs achieved power conversion efficiencies (PCEs) of 21.46% for small-area devices (0.108 $cm^2$) and 19.38% for large-area modules (15.52 $cm^2$), representing the highest reported values for fully vacuum-deposited devices fabricated using eSAM. The unencapsulated device exhibited excellent operational stability maintained 91.9% of its initial efficiency following 1200 hours of continuous maximum power point tracking (MPPT) operation under $N_2$ atmosphere with LED light soaking. Furthermore, for devices incorporating solution-processed perovskite, a champion efficiency of 23.67% was achieved, which also stands as the highest performance reported to date for this system utilizing eSAM. This work elucidates thickness-dependent transport mechanisms for eSAM, advancing the in-depth application of vacuum deposition technology in perovskite photovoltaics.

## RESULTS and DISCUSSION

The reproducibility of devices fabricated based on eSAM is significantly higher than that of devices prepared using solution-processed SAM (sSAM), which contributes to the commercialization of PSCs (Figures S1 and S2). Figure 1 systematically illustrates the thickness regulation strategy of Me-4PACz layers and their integrated application in fully vacuum-evaporated PSCs and solution-processed PSCs. The SAM material Me-4PACz is initially deposited on ITO substrates through evaporation, followed by the formation of perovskite films via a sequential two-step evaporation process or the solution process. The evaporated Me-4PACz layer resides between the ITO layer and perovskite film. For solution-processed Me-4PACz layer, it demonstrates superior charge transport capabilities through its ultra-thin monolayer configuration.[25,26] However, the aggregation tendency of Me-4PACz materials leads to uneven film coverage, which significantly degrades the performance of the resulting perovskite photovoltaic devices.[27] Similarly, thin evaporated Me- 4PACz layers (Thin eSAM) still exhibit intermolecular aggregation and poor coverage. In contrast, thick Me-4PACz layer (Acting as Thick eSAM) significantly enhances molecular coverage at both the ITO surface and perovskite buried surface, thereby facilitating stable hole transport. Regarding molecular alignment, for the Thin eSAM, the specific anchoring of phosphonic acid groups to the ITO substrate induces a vertically aligned molecular orientation. With increasing thickness, while maintaining vertical alignment in bottom layers, upper molecules transition to horizontal orientation, forming a distinctive vertical-to-horizontal composite alignment structure. The horizontally aligned molecular layers exhibit more negative highest occupied molecular orbital (HOMO) energy levels compared to vertically oriented counterparts (but still higher than the perovskite HOMO energy level), establishing an energy-level gradient barrier tha optimizes hole transport at the buried interface.

To elucidate the coverage characteristics of thin versus thick Me-4PACz layer between ITO and perovskite layers, molecular dynamics simulations were employed. As shown in Figure 2a, we made a stack of ITO/Me-4PACz/perovskite and performed molecular dynamics simulations for 20 ns. The specific parameter settings of the stacked model are shown in Figure S3. Two cases are considered, where a very thin one layer of Me-4PACz (representing Thin eSAM) and a thicker two layer Me- 4PACz (representing Thick eSAM). For clearer visualization of Me-4PACz distribution, the ITO layers are made invisible. It is observed that Thick eSAM makes a better surface coverage. In addition, we quantified the surface coverage ability of Me-4PACz molecules from "surface non-uniformity" (SNU), calculated from the standard deviation of number of molecules on a surface grid (Figure S4). Two example diagrams showing a good surface coverage quantified by a high SNU and a bad surface coverage by a low SNU (Figure S5). The time-dependence of SNU was calculated for Thin eSAM and Thick eSAM, and it was found that Thick eSAM has a high SNU, indicating a uniform surface coverage.

For precise observation of Me-4PACz molecular coverage distribution at the ITO surface and the perovskite buried surface, we calculated the radial-pair distribution function (RDF) in ITO/ Me- 4APCz/perovskite stack for Thin eSAM and Thick eSAM. Figure 2b (top) represents the RDF between the Me-4PACz molecules and the underlying perovskite surface. The Thin eSAM shows higher peaks, indicating a stronger, but potentially more localized, interaction between the Me- 4PACz molecules and the perovskite buried surface. The Thick eSAM exhibits lower, broader peaks, suggesting a more distributed and uniform interaction between the SAM and the perovskite buried surface. Consequently, the Thick eSAM significantly improved uniform coverage of Me-4PACz molecules at the perovskite buried interface. Figure 2b (bottom) represents the RDF values as a function of distance (in Å) for the surface-adsorbed monolayers (Me-4APCz) on the ITO substrate. We see that, the Thin eSAM exhibits sharper and higher peaks, suggesting localized clustering and less uniform arrangement within the Me-4PACz. However, the Thick eSAM shows broader and smoother peaks, indicating more uniform molecular arrangement and better coverage across the surface.

Additionally, X-ray photoelectron spectroscopy (XPS) tests were conducted to further investigate the variation in molecular coverage on the ITO surface induced by Me-4PACz films of different thicknesses. Firstly, by testing samples for eSAM on Cu and ITO substrates, the P 2*p* peak shifts toward higher binding energy (Figure S6), indicating that the phosphate groups in the SAM interact with ITO to form In-O-P bonds. The O 1*s* spectra of Thin eSAM and Thick eSAM on ITO are shown in Figures 2c and 2d, respectively, with the original test curves presented in Figures S7 and S8. The oxygen peaks can be deconvoluted into four components: In-O-P/In-O-H peak, In/Sn-O peak, surface $V_O$ peak, and $H_2O$ peak.[28] The In-O-P/In-O-H peak area ratio for the Thick eSAM is 23.8%, significantly higher than that of the Thin eSAM (16.3%). This further confirms that the Thick eSAM enhances the molecular coverage of Me-4PACz on the ITO surface, which aligns with the RDF calculation results. Furthermore, the coverage uniformity of SAM can be visually observed by conductive atomic force microscopy (C-AFM), owing to conductivity of ITO being two orders of magnitude higher than SAM when used as an electrode. The C-AFM test results for thin and thick eSAM on ITO are shown in Figure S9. The exported diagonal line scan figure (Figure 2e) reveals the presence of local speckled current spikes within the thin eSAM. These spikes are attributed to the detection of the underlying ITO substrate due to poor molecular coverage. In contrast, for the thick eSAM, such speckled current spikes completely disappear, indicating uniform coverage of the SAM film. In summary, the thicker Me-4PACz layer promotes more uniform molecular coverage, thereby enabling stable and efficient hole transport in PSCs.

For Thick eSAM, it is observed that increased thickness not only enhances surface coverage but also forms a vertical-to-horizontal molecular orientation, thereby improving hole transport properties. Molecular dynamics simulations revealed distinct molecular orientation changes in layered structures. As shown in Figure 3a, two vertical Me-4PACz layers of initial structures were modeled (one anchored to ITO and the upper layer remaining free). The final simulation structure demonstrates that the anchored layer maintains near-vertical orientation, while the free upper layer adopts a horizontal orientation, confirming the vertical-to-horizontal orientation in Thick eSAM. The current density-voltage (*J-V*) curves of fully evaporated PSCs with different Me-4PACz thicknesses are presented in Figure S10, with corresponding PCE statistics in Figure S11. Devices employing 20 nm Me-4PACz showed optimal performance, subsequently, Thick eSAM (20 nm) will be compared with Thin eSAM (2 nm) for comprehensive analysis. The angle-dependent p-polarized-resolved PL measurements provided quantitative orientation analysis through order parameter ($S$) determination. As illustrated in Figure S12, $S = 1$ indicates that molecules are perfectly vertical orientation for the substrate surface, $S = 0$ means random orientation of molecules, and $S = -0.5$ indicates that molecules are perfectly horizontal orientation for the substrate surface.[29] Angle-dependent PL spectra are shown in Figure 3b, the $S$ value for 2 nm Thin eSAM is 0.206 (within the 0–1 range), indicating partial vertical molecular orientation. The incomplete vertical orientation may stem from its non-monolayer structure. In contrast, the $S$ value for 20 nm Thick eSAM is -0.478, approaching the theoretical horizontal limit ($S = -0.5$), which confirms a predominantly horizontal molecular alignment parallel to the substrate surface. Therefore, this confirms the vertical-to-horizontal transition observed in simulations. In addition, water contact angle measurements further supported orientation differences (Figure 3c). The Thin eSAM displayed a large contact angle due to exposed hydrophobic carbazole/methyl groups by molecule vertical orientation, while the Thick eSAM showed a significantly reduced angle from exposed hydrophilic phosphate groups by molecule horizontal alignment. Further, compared to Thin eSAM, the smaller difference for the left and right contact angle in Thick eSAM suggests improved uniformity, this echoes the results of the previous discussion.

To further investigate the effects of Thin eSAM and Thick eSAM on hole transport, ultraviolet photoelectron spectroscopy (UPS) was employed to examine changes in the film's energy band structure. As shown in Figure 3d, UPS measurements were conducted for Thin eSAM and Thick eSAM, with their valence bands calculated to be -5.78 eV and -5.82 eV, respectively. The primary reason for this difference is the degree of impact for the interfacial dipole moment generated by the interaction between SAM molecules and the ITO substrate. Thick eSAM exhibits a lower HOMO energy level due to its weaker response to this interfacial dipole moment. Correspondingly, UPS measurement for perovskite layer was also measured, revealing a valence band of -5.85 eV (Figure S13). As a result, for Thick eSAM, the horizontally oriented Me-4PACz molecules exhibit a more negative HOMO energy level compared to vertically oriented ones (not exceed that of the perovskite layer), the spontaneous formation of the graded barrier facilitates hole extraction and transport (Figure 3e). In addition, atomic force microscopy (AFM) was performed to analyze the surface roughness of Thin eSAM and Thick eSAM, assessing their impact on subsequent perovskite film growth. The Thick eSAM exhibits a significantly lower root mean square (RMS) roughness of 0.49 nm compared to the Thin eSAM (2.41 nm), indicating a smoother surface (Figure 3f). The bearing area ratio curve in Figure 3f further highlights the height variations in the Me-4PACz film morphology. The Thick eSAM shows a steeper curve, suggesting smaller height fluctuations and higher uniformity, which is more favorable for perovskite film growth.

In conclusion, a schematic of the molecular arrangement in the Thick eSAM between ITO and the perovskite layer is depicted in Figure 3g. The bottom molecules anchor vertically on the ITO surface, while the upper molecules adopt a horizontal orientation, resulting in a vertical-to-horizontal molecular alignment. Moreover, the increased film thickness enhances ITO surface coverage by the bottom molecules, while the exposed phosphonic acid groups in the top layers act as passivating moieties, can effectively reducing defects at the perovskite buried interface.

To evaluate the influence of Me-4PACz layer thickness on perovskite film growth, we first employed back grazing-incidence wide-angle X-ray scattering (Back-GIWAXS) technique to probe the quality of the perovskite buried interface (Figure 4a). The Back-GIWAXS technique enables direct probing of perovskite signals at the buried interface without exfoliation, thereby avoiding potential test errors caused by film damage during peeling. Figure S14 displays Back-GIWAXS patterns of ITO/Me-4PACz/perovskite samples (on flexible substrates) prepared with 2 nm and 20 nm Me-4PACz. The corresponding integrated Back-GIWAXS (Figure 4b) reveals that under comparable background peak intensities, the (100) diffraction peak intensity of the perovskite film based on Thick eSAM is significantly stronger than the perovskite film based on Thin eSAM, this indicates that Thick eSAM promotes favorable perovskite crystallization.[30] Additionally, PL mapping (Figure 4c) demonstrates improved film uniformity for the perovskite layer based on Thick eSAM, this may be attributed to the reduced surface roughness of Thick eSAM. The enhancement in the flatness of Me-4PACz film promotes homogeneous perovskite growth and improves perovskite quality.

As previously mentioned, the horizontal orientation of Me-4PACz exposes phosphate groups, thereby providing passivating functional groups. Fourier-transform infrared (FTIR) analysis confirms interactions between P=O groups in phosphate groups and $Pb^{2+}$ ions (Figures S15 and S16), which can passivate undercoordinated $Pb^{2+}$ ions at the buried interface. Drive-level capacitance profiling (DLCP) measurement (Figure 4d) of devices with Thin eSAM and Thick eSAM reveal overlapping left-side curves, indicating negligible influence of thickness for Me-4PACz film on upper interface defects. However, the right-side DLCP curve of the device Thick eSAM with lies markedly lower than the Thin eSAM, confirming that thicker Me-4PACz effectively suppresses buried interface defects, this further suggests that defects at the buried interface are passivated more effectively. The space-charge limited current (SCLC) testing further supports this conclusion. Hole-only devices (ITO/Me-4PACz/Perovskite/Me-4PACz/Ag) with 20 nm bottom Me-4PACz exhibit a lower trap-filling limit voltage ($V_{TFL}$ = 0.43 V) compared to 2 nm devices ($V_{TFL}$ = 0.68 V) (Figure 4e). Notably, their upper Me-4PACz layers have identical thickness. Since defect density is proportional to $V_{TFL}$,[31,32] this confirms reduced trap states in perovskite films grown on thicker Me- 4PACz. In addition, dark current measurements (Figure S17) further corroborate this finding, showing lower dark current densities in devices with Thick eSAM. The reduction in defects leads to decreased non-radiative recombination, resulting in lower dark current density for device with Thick eSAM. Collectively, these results demonstrate that Thick eSAM enhances interfacial passivation, leading to reduced defect density and improved device performance.

To further assess the influence of Me-4PACz thickness on hole transport, systematic characterizations were performed. First, photocurrent density-effective voltage ($J_{ph}$-$V_{eff}$) analysis (Figure 4f) demonstrates enhanced charge collection and exciton dissociation in devices with thicker Me-4PACz film. $J_{ph}$ is calculated as the subtraction of $J_d$ from $J_L$, with $J_L$ corresponding to the current density of the perovskite device under 100 mW cm$^{-2}$ illumination and $J_d$ representing the dark-state current density. $V_{eff}$ is derived by offsetting the compensation voltage ($V_0$) against the applied bias voltage ($V_{bias}$), where $V_0$ specifically indicates the voltage at which $J_{ph}$ becomes zero. By increasing effective voltage, devices fabricated with Thick eSAM reaches saturated photocurrent ($J_{sat}$) earlier than the device with Thin eSAM. This indicates improved charge collection and extraction efficiency, further demonstrating that the use of Thick eSAM enhances hole transport properties.[33,34] Furthermore, photoluminescence (PL) measurements (Figure 4g) were performed with the test sample structure being ITO/ Me-4PACz/Perovskite. As shown in Figure 4g, the perovskite film based on Thick eSAM exhibited significantly reduced PL intensity, consistent with the PL mapping results, indicating improved hole transport. Time-resolved photoluminescence

(TRPL) measurements further revealed a decreased average carrier lifetime of 642.2 ns for the perovskite film based on Thick eSAM, substantially lower than the 877.6 ns observed for the perovskite film based on Thin eSAM (Figure 4h). This further confirms the enhanced hole transport and demonstrates that the use of Thick eSAM leads to significantly reduced non-radiative recombination at the bottom interface. The TRPL fitting parameters are summarized in Table S1. In conclusion, these measurements demonstrate improved hole transport and extraction, which may be attributed to synergistic effects of gradient energy barrier formation, defect passivation, and enhanced perovskite quality enabled by Thick eSAM.

Leveraging the pronounced advantages of Thick eSAM over its thin film counterpart, fully vacuum-deposited PSCs based on Thick eSAM exhibited substantially enhanced device performance. The device architecture follows ITO/Me-4PACz/Perovskite/C60/BCP/Ag. Figure 5a displays the $J$-$V$ characteristic curves of small-area devices (active area: 0.108 cm$^2$) employing 2 nm and 20 nm Me- 4PACz films, with detailed parameters summarized in Table S2. The device based on Thick eSAM achieved a remarkable PCE improvement from 18.48% to 21.46%, accompanied by concurrent enhancements in short-circuit current density ($J_{sc}$), open-circuit voltage ($V_{oc}$), and fill factor (FF). Reduced hysteresis in the Thick eSAM-based device was further confirmed by forward and reverse scans of $J$-$V$ curves (Figure S18). These improvements are attributed to the Thick eSAM facilitating superior perovskite quality and efficient hole transport. External quantum efficiency (EQE) measurements (Figure 5b) corroborated the $J_{sc}$ improvement, with the device with Thick eSAM exhibiting an EQE-derived $J_{sc}$ of 24.37 mA cm$^{-2}$, consistent with $J$-$V$ results of devices (<3% discrepancy).

To validate scalability, large-area modules (25 cm$^2$ substrate, 15.52 cm$^2$ active area) were fabricated, demonstrating a PCE increase from 18.02% to 19.38% (Figure 5c) with consistent $J_{sc}$, $V_{oc}$, and FF enhancements (Table S3). This underscores the universal applicability of Thick eSAM for performance optimization. To evaluate the versatility of evaporated Me-4PACz, we integrated it into highly-efficient solution-processed inverted PSCs, device structure is ITO/Me- 4PACz/Perovskite/PCBM/BCP/Ag. As shown in Figure 5d and Table S4, devices with evaporated thick eSAM achieved a remarkable PCE of 23.67% which demonstrated great potential. Correspondingly, devices using thick eSAM film also exhibited a lower hysteresis effect (Figure

S19). Furthermore, the stable PCE was measured over 3 min at the maximum power output point for the fully vacuum-deposited devices and solution-processed devices. For the vacuum-deposited devices (Figure S20), the average steady-state output PCE of the thick eSAM-based devices is 21.25%, which is higher than that of the thin eSAM-based devices (18.20%). And for the solution-processed devices (Figure S21), the average steady-state output PCE of the thick eSAM-based devices reaches 23.48%, which is also higher than that of the thin eSAM-based devices (21.72%). Both of these results are in good agreement with the data obtained from the *J-V* curve measurements. A comprehensive review was conducted of all currently available fully vacuum-deposited PSCs based on eSAM as well as solution-processed PSCs (Figure 5e). In our work, by utilizing Thick eSAM, impressive efficiencies were achieved in both types of device architectures. Device batch-to-batch reproducibility, a critical metric for industrial feasibility,[35-38] was systematically evaluated. The PCE variance ($S^2$) of fully vacuum-deposited PSCs (target) and solution-processed PSCs (control) was compared (Figure 5f). For 20 devices within the same batch (Figure S22), the fully vacuum-deposited PSCs exhibit a lower $S^2$ of 0.11 for PCE versus 1.85 for solution-processed devices. Similarly, across 10 independent batches (Figure S23), the fully vacuum-deposited PSCs show a $S^2$ of 0.23 for PCE, significantly lower than the solution-processed counterparts ($S^2$ = 3.06). The statistical distribution plots of their $J_{sc}$, $V_{oc}$, and FF device parameters are shown in Figures S24 and S25, with the fully vacuum-deposited PSCs exhibiting greater stability in all parameters. These results underscore the superior reproducibility and stability of the fully vacuum-deposited fabrication process. Finally, to further investigate the impact of Thin eSAM and Thick eSAM on device performance, stability tests were conducted. The stability-test devices were fabricated using the aforementioned method with a structure of ITO/Me-4PACz/Perovskite/C60/BCP/Ag. As shown in Figure 5g, the stability under the ISOS-D-1 protocol was first evaluated. Unencapsulated devices were stored under ambient conditions (a relative humidity of 30-40%, ~25°C) in the dark and subjected to periodic PCE measurements. Over time, the device with thick eSAM retained 91.5% of its initial PCE after 1,176 hours, whereas the device based on thin eSAM maintained only 66.9% of its initial PCE under identical conditions, demonstrating the superior storage stability of device with Thick eSAM. Furthermore, as illustrated in Figure 5h, the devices with Thin eSAM and Thick eSAM underwent maximum power point tracking (MPPT) measurement under continuous illumination in an $N_2$ atmosphere using a white LED (1-sun equivalent intensity, ~35°C). After 1,200 hours of MPPT operation, the device with Thick eSAM retained 91.9% of its initial PCE, significantly outperforming the device with Thin eSAM (70.0% retention). This enhancement is likely attributed to the improved perovskite quality and efficient hole transport enabled by Thick eSAM. These results further highlight the critical role of Thick eSAM in enhancing long-term device stability.

## Conclusion

In summary, we have elucidated the critical role of eSAM thickness in governing interfacial dynamics and hole transport efficiency. By systematically investigating, we have demonstrated that Thick eSAM spontaneously forms a vertical-to-horizontal heterogeneous molecular orientation, which establishes gradient energy-level alignment. This unique configuration not only optimizes hole extraction via graded energy barriers but also suppresses interfacial recombination through defect passivation. Additionally, Thick eSAM enhances bilateral coverage at both the ITO and perovskite buried surface, further improving hole transport. The enhanced interfacial quality and charge transport dynamics directly translate to impressive PCE of 21.46% and 19.38% in small-area devices and large-area modules for fully vacuum-deposited devices fabricated using eSAM, respectively. Furthermore, combining eSAM with solution-processed perovskite achieved a champion efficiency of 23.67%, setting an impressive value for this system using eSAM. Our findings resolve ambiguities regarding the thickness-dependent transport mechanism of eSAM and unlock eSAM potential for optimizing scalable, solvent-free photovoltaic manufacturing. In the future, structural modifications of the SAM molecules, including alterations to the alkyl chain length, adjustments to the terminal groups and changes to intermolecular interactions, can be pursued to better advance this strategy. By optimizing molecular orientation and stacking, the advantages of eSAM can be further enhanced, ultimately leading to improved performance in perovskite photovoltaics.

## EXPERIMENTAL PROCEDURES

**Materials and Solutions**

Dimethyl sulfoxide (DMSO; anhydrous, 99.9%), N,N-dimethylformamide (DMF; anhydrous, 99.8%), Chlorobenzene (CB; anhydrous, 99.8%), Ethanol, Formamidinium iodide (FAI; 99%), and Silver (Ag) were all purchased from Sigma-Aldrich Inc. Indium in oxide (ITO) substrates, Methylammonium chloride (MACl; 99%), (4-(3,6-dimethyl-9H-carbazol-9-yl) butyl) phosphonic acid (Me-4PACz), Bathocuproine (BCP) and C60 were all purchased from Advanced Election Technology Co., Ltd. Lead chloride ($PbCl_2$), Cesium iodide (CsI), and Lead iodide ($PbI_2$) were all purchased from Xi'an Yuri Solar Co., Ltd. Phenyl-C61-butyric acid methyl ester (PCBM) was purchased from 1-Material. All the materials were used as received without further purification. For the perovskite precursor solution of composition $FA_{0.95}Cs_{0.05}PbI_3$, 1.5 M perovskite precursor solution was prepared by mixing CsI, FAI and $PbI_2$ in DMF:DMSO = 4:1 (v/v) mixed solvent subject to the stochiometric ratio. In addition, 20 mg PCBM was dissolved in 1 mL of CB to prepare the electron transport layer solution.

**Device Fabrication**

For the fully vacuum-deposited perovskite solar cells, the PSCs have a structure as ITO/Me-4PACz/perovskite/C60/BCP/Ag. The ITO substrates were sequentially cleaned in an ultrasonic bath for 15 minutes each with a glass cleaning solution, deionized water, acetone and ethanol. Different thickness of Me-4PACz film was deposited on ITO at a rate of 0.4 Å s$^{-1}$. $PbI_2$, $PbCl_2$, and CsI were co-evaporated at the rate of 7 Å s$^{-1}$, 0.5 Å s$^{-1}$, and 0.3 Å s$^{-1}$, respectively, until the total film thickness was 300 nm. Then, FAI was evaporated at a rate of 2 Å s$^{-1}$ for 350 nm. Finally, it was annealed at 150 °C for 15 minutes to finish the preparation of the perovskite films. Subsequently, a 20 nm C60 layer was deposited at a rate of 0.50 Å s$^{-1}$, followed by an 8 nm BCP layer at the same deposition rate, and then Ag was deposited by thermal evaporation at a rate of 2 Å s$^{-1}$ for 80 nm. All depositions were carried out under a vacuum pressure of $5 \times 10^{-4}$ Pa. The active areas for small-area devices and large-area modules are 0.108 cm$^2$ and 15.52 cm$^2$, respectively. The effective area of the mask used in the testing process is 0.0669 cm$^2$.

For solution-processed perovskite solar cells, PSCs were fabricated with the following structure: ITO/Me-4PACz/perovskite/PCBM/BCP/Ag. The cleaning procedure for the ITO substrate is the same as mentioned above. For the evaporated Me-4PACz layer, it was deposited on substrates at a rate of 0.4 Å s$^{-1}$. After that, the perovskite film was obtained by spin-coating at 1000 rpm for 10 s and 6000 rpm for 35 s. During the spin-coating, 250 μL CB solution was dripped at 15 s before ending. The film was then annealed at 100 ℃ for 40 min. Then, PCBM solution was spin-coated on the perovskite film at 2000 rpm for 40 s. Finally, BCP (10 nm) and Ag (120 nm) were sequentially evaporated in a thermal evaporation chamber onto the PCBM film under a vacuum pressure of 5 × 10$^{-4}$ Pa.

## Measurements

The XPS and UPS spectra were obtained through a Kratos AXIS Ultra-DLD ultrahigh vacuum surface analysis system. Angle-dependent PL spectra were obtained by a Hamamatsu's established molecular orientation measurement system(C13472-01, Hamamatsu Photonics). The AFM images were measured through contact mode by a MultiMode V atomic force microscopy (Veeco, USA). The Back-GIWAXS patterns were measured at the BL14B1 beamline of the Shanghai Synchrotron Radiation Facility using X-ray with a wavelength of 1.24 Å. The PL mapping spectra were obtained by the excitation wavelength of 365 nm through a commercial confocal Raman/PL imaging system (WITEC Alpha300R). The DLCP, dark $J$-$V$ curves were obtained by a Keithley 4200 digital source meter co-operating with a Lake Shore Cryotronics probe stage in the dark. The SCLC curves were measured after removing the lamp. The $J_{ph}$-$V_{eff}$ curves were obtained using a programmable Keithley 2400 source meter. The solar cells obtained the $J$-$V$ curve by irradiated under AM 1.5G solar (100 mW/cm$^2$) using a programmable Keithley 2400 source meter. A system combinate by xenon lamp, monochromator, chopper, lock-in amplifier and calibrated silicon photodetector was used to measure the EQE. The MPPT was tested by using the LED light (WAVELABS LS-2). PL spectra were measured through Horiba Jobin-Yvon LabRAM HR800. TRPL decay was recorded through a transient fluorescence spectrometer (HORIB-FM-2015). FTIR spectroscopy measurements were conducted by using absorption infrared spectrometer (VERTX 70).

## SUPPLEMENTAL INFORMATION

## ACKNOWLEDGMENTS

The authors acknowledge financial support from the National Natural Science Foundation of China (No. 52273189), the Natural Science Foundation of Jiangsu Province (Nos. BG2024016, BZ2023052, BE2022026-2, BK20240756), the Natural Science Foundation of Anhui Province (No. 202423h08050004), the China Postdoctoral Science Foundation (Nos. 2024T170622, 2023M742526, GZB20240518), the Jiangsu Funding Program for Excellent Postdoctoral Talent (No. 2024ZB061), the Suzhou science and technology plan project (Nos. ST202212, ST202312). This work is also supported by Suzhou Key Laboratory of Functional Nano & Soft Materials. Collaborative Innovation Center of Suzhou Nano Science & Technology, the 111 Project, Joint International Research Laboratory of Carbon-Based Functional Materials and Devices.

## AUTHOR CONTRIBUTIONS

L.H., K.L.W. and Z.K.W. conceived the idea. K.L.W. and Z.K.W supervised the project. I.Y. provided software support. L.H. designed the experimental approach and drafted the manuscript. Z.C., Z.S., Z.H.S., S.M., X.C., J.C., C.H.C., Y.X., Y.T.Y., J.C.L., D.N. contributed to the execution of experiments and characterization. L.H., I.Y. and Z.K.W. refined and polished the final version. All authors reviewed and approved the manuscript.

## REFERENCES

1.  Zhao, K., Liu, Q., Yao, L., Değer, C., Shen, J., Zhang, X., Shi, P., Tian, Y.; Luo, Y., Xu, J., et al. (2024). peri-Fused polyaromatic molecular contacts for perovskite solar cells. Nature *632*, 301-306, https://doi.org/10.1038/s41586-024-07712-6.

2.  Wang, K., Li, X., Lou, Y., Li, M., and Wang, Z. (2021). $CsPbBrI_2$ perovskites with low energy loss for high-performance indoor and outdoor photovoltaics. Sci. Bull. *66*, 347-353, https://doi.org/10.1016/j.scib.2020.09.017.

3.  Chen, J., Deger, C., Su, Z., Wang, K., Zhu, G., Wu, J., He, B., Chen, C., Wang, T. Gao, X., et al. (2024). Magnetic-biased chiral molecules enabling highly oriented photovoltaic perovskites. Natl. Sci. Rev. *11*, nwad305, https://doi.org/10.1093/nsr/nwad305.

4.  Huang, Z., Bai, Y., Huang, X., Li, J., Wu, Y., Chen, Y., Li, K., Niu, X., Li, N., Liu, G., et al. (2023). Anion–π interactions suppress phase impurities in $FAPbI_3$ solar cells. Nature *623*, 531-

537, https://doi.org/10.1038/s41586-023-06637-w.

5. Wang, K., Yang, Y., Lou, Y., Li, M., Igbari, F., Cao, J., Chen, J., Yang, W., Dong, C., Li, L., et al. (2021). Smelting recrystallization of CsPbBrI$_2$ perovskites for indoor and outdoor photovoltaics. eScience *1*, 53-59, https://doi.org/10.1016/j.esci.2021.09.001.

6. Sutanto, A. A., Caprioglio, P., Drigo, N., Hofstetter, Y. J., Garcia-Benito, I., Queloz, V. I. E., Neher, D., Nazeeruddin, M. K., Stolterfoht, M., Vaynzof, Y., et al. (2021). 2D/3D perovskite engineering eliminates interfacial recombination losses in hybrid perovskite solar cells. Chem. *7*, 1903-1916, https://doi.org/10.1016/j.chempr.2021.04.002.

7. Chen, C., Liu, G., Chen, X., Deger, C., Jin, R., Wang, K., Chen, J., Xia, Y., Huang, L., Yavuz, I., et al. (2025). Methylthio substituent in SAM constructing regulatory bridge with photovoltaic perovskites. Angew. Chem. Int. Ed. *64*, e202419375, https://doi.org/10.1002/anie.202419375.

8. Tang, H., Shen, Z., Shen, Y., Yan, G., Wang, Y., Han, Q., and Han, L. (2024). Reinforcing self-assembly of hole transport molecules for stable inverted perovskite solar cells. Science *383*, 1236-1240, https://doi.org/10.1126/science.adj9602.

9. Zhou, J., Luo, Y., Li, R., Tian, L., Zhao, K., Shen, J., Jin, D., Peng, Z., Yao, L., Zhang, L., et al. (2025). Molecular contacts with an orthogonal π-skeleton induce amorphization to enhance perovskite solar cell performance. Nat. Chem. *17*, 564-570, https://doi.org/10.1038/s41557-025-01732-z.

10. Wu, H., Wu, J., Zhang, Z., Guan, X., Wang, L., Deng, L., Li, G., Abate, A., and Li, M. (2025). Tailored Lattice-Matched Carbazole Self-Assembled Molecule for Efficient and Stable Perovskite Solar Cells. J. Am. Chem. Soc. *147*, 8004-8011, https://doi.org/10.1021/jacs.5c00629.

11. Nematov, D. D., Kholmurodov, Kh. T., Yuldasheva, D. A., Rakhmonov, Kh. R., and Khojakhonov, I. T. (2022). Ab-initio Study of Structural and Electronic Properties of Perovskite Nanocrystals of the CsSn[Br$_{1-x}$I$_x$]$_3$ Family. HighTech and Innovation Journal, 3(2), 140–150. https://doi.org/10.28991/HIJ-2022-03-02-03

12. Ouedraogo, N., Ouyang, Y., Guo, B., Xiao, Z., Zuo, C., Chen, K., He, Z., Odunmbaku, G., Ma, Z., Long, W., et al. (2024). Printing Perovskite Solar Cells in Ambient Air: A Review. Adv. Energy Mater. *14*, 2401463, https://doi.org/10.1126/science.aaf8060.

13. Nematov, D., Kholmurodov, K., Stanchik, A., Fayzullaev, K., Gnatovskaya, V., and Kudzoev, T. (2023). On the Optical Properties of the Cu2ZnSn [S$_{1-x}$Se$_x$]4 System in the IR Range. Trends in Sciences, 20(2), 4058. https://doi.org/10.48048/tis.2023.4058

14. Fan, B., Xiong, J., Zhang, Y., Gong, C., Li, F., Meng, X., Hu, X., Yuan, Z., Wang, F., and Chen, Y. (2022). A bionic interface to suppress the coffee-ring effect for reliable and flexible perovskite modules with a near-90% yield rate. Adv. Mater. *34*, 2201840, https://doi.org/10.1002/adma.202201840.

15. Nematov, D. D. (2021). Investigation optical properties of the orthorhombic system CsSnBr3-xIx: application for solar cells and optoelectronic devices. Journal of Human, Earth, and Future, 2(4), 404–411. https://doi.org/10.28991/HEF-2021-02-04-08

16. Farag, A., Feeney, T., Hossain, I. M., Schackmar, F., Fassl, P., Küster, K., Bäuerle, R., Ruiz-Preciado, M. A., Hentschel, M., Ritzer, D. B., et al. (2023). Evaporated self-assembled monolayer hole transport layers: lossless interfaces in p-i-n perovskite solar cells. Adv. Energy Mater. *13*, 2203982, https://doi.org/10.1002/aenm.202203982.

17. Jiang, W., Wang, D., Shang, W., Li, Y., Zeng, J., Zhu, P., Zhang, B., Mei, L., Chen, X., Xu, Z., et al. (2024). Spin-Coated and Vacuum-Processed Hole-Extracting Self-Assembled Multilayers with H-Aggregation for High-Performance Inverted Perovskite Solar Cells. Angew. Chem. Int. Ed. *63*, e202411730, https://doi.org/10.1002/anie.202411730.

18. Park, S. J., Hong, G. P., and Kim, J. Y. (2025). Vacuum-Deposited Self-Assembled Monolayers for Perovskite/PERC Tandem Solar Cells. ACS Energy Lett. *10*, 3743-3745, https://doi.org/10.1021/acsenergylett.5c01338.

19. Zheng, Y., Zhan, Z., Pang, N., Lu, Y., Lin, Z., Shi, T., Chen, K., Lin, D., Jiang, Y., and Xie, W., (2025). Gradient Doping for Stress-Relief in Vapor-deposited Perovskite Film to Achieve High-performance p-i-n Perovskite Solar Cells with a 23% Efficiency. Adv. Mater. *37*, 2501162, https://doi.org/10.1002/adma.202501162.

20. Vollbrecht, J., Barnscheidt, V., Clausing, R., Löhr, J., Mettner, L., Neuba, A., Raugewitz, A., Strey, J., and Peibst, R. (2025). Less is more: Enabling Solvent-Free Fabrication of Perovskite Solar Cells via Thermal Evaporation of Ultrathin Self-Assembled Monolayers. Sol. RRL, *0*, e70087, https://doi.org/10.1002/solr.202500429.

21. Nematov, D. (2022). DFT calculations of the main optical constants of the Cu2ZnSnSexS4-x system as high-efficiency potential candidates for solar cells. International Journal of Applied Power Engineering (IJAPE), 11(4), 287–293. https://doi.org/10.11591/ijape.v11.i4.pp287-293

22. Nematov, D. (2022). Influence of Iodine Doping on the Structural and Electronic Properties of CsSnBr3. International Journal of Applied Physics, 7, 36–47.

23. Gil-Escrig, L., Momblona, C., La-Placa, M. G., Boix, P. P., Sessolo, M., and Bolink, H. J. (2018). Vacuum deposited triple-cation mixed-halide perovskite solar cells. *Adv. Energy Mater.*, *8*, 1703506, https://doi.org/10.1002/aenm.201703506.

24. Piot, M., Alonso, J. E. S., Zanoni, K. P., Rodkey, N., Ventosinos, F., Roldán-Carmona, C., Sessolo, M. and Bolink, H. (2023). Fast Coevaporation of 1 μm Thick Perovskite Solar Cells. *ACS Energy Lett.*, *8*, 4711-4713, https://doi.org/10.1021/acsenergylett.3c01724.

25. Yu, X. Y., Sun, X. L., Zhu, Z. L., and Li, Z. A. (2025). Stabilization Strategies of Buried Interface for Efficient SAM-Based Inverted Perovskite Solar Cells. Angew. Chem. Int. Ed. *64*, e202419608, https://doi.org/10.1002/anie.202419608.

26. Nematov, D., and Hojamberdiev, M. (2025). Machine Learning-Driven Materials Discovery: Unlocking Next-Generation Functional Materials – A review. Computational Condensed Matter, 45, e01139. https://doi.org/10.1016/j.cocom.2025.e01139

27. Li, M. L., Liu, M., Qi, F., Lin, F. R., and Jen, A. K. Y. (2024). Self-assembled monolayers for interfacial engineering in solution-processed thin-film electronic devices: design, fabrication, and applications. Chem. Rev. *124*, 2138-2204, https://doi.org/10.1021/acs.chemrev.3c00396.

28. .Nematov, D. D., Burhonzoda, A. S., Kholmurodov, K. T., Lyubchyk, A. I., and Lyubchyk, S. I. (2023). A Detailed Comparative Analysis of the Structural Stability and Electron-Phonon Properties of ZrO2: Mechanisms of Water Adsorption on t-ZrO2 (101) and t-YSZ (101) Surfaces. Nanomaterials, 13(19), 2657. https://doi.org/10.3390/nano13192657

29. Yokoyama, D. (2011). Molecular orientation in small-molecule organic light-emitting diodes. *J. Mater. Chem. 21*, 19187-19202, https://doi.org/10.1039/C1JM13417E.

30. Bian, Z. K., Su, Z. H., Lou, Y. H., Chen, J., Jin, R. J., Chen, C. H., Xia, Y., Huang, L., Wang, K. L., Gao, X. Y., et al. (2025). Removal of Residual Additive Enabling Perfect Crystallization of Photovoltaic Perovskites. Angew. Chem. Int. Ed. *64*, e202416887, https://doi.org/10.1002/anie.202416887.

31. Yang, Y. T., Hu, F., Teng, T. Y., Chen, C. H., Chen, J., Nizamani, N., Wang, K. L., Xia, Y., Huang, L., and Wang, Z. K. (2025). Dual-Stage Reduction Strategy of Tin Perovskite Enables High Performance Photovoltaics. Angew. Chem. Int. Ed. *137*, e202415681, https://doi.org/10.1002/anie.202415681.

32. Nematov, D. (2024). Analysis of the Optical Properties and Electronic Structure of Semiconductors of the Cu2NiXS4 (X = Si, Ge, Sn) Family as New Promising Materials for Optoelectronic Devices. Journal of Optics and Photonics Research, 1(2), 91–97. https://doi.org/10.47852/bonviewJOPR42021819

33. Chen, X., Chen, C. H., Su, Z. H., Chen, J., Wang, K. L., Xia, Y., Nizamani, N., Huang, L., Jin, R. J., Li, Y. H., et al. (2025). Adhesively Bridging SAM Molecules and Perovskites for Highly Efficient Photovoltaics. Adv. Funct. Mater. *35*, 2415004, https://doi.org/10.1002/adfm.202415004.

34. Chen, C. H., Hu, F., Wang, K. L., Chen, J., Teng, T. Y., Shi, Y. R., Xia, Y., Su, Z. H., Gao, X.

Y., Yavuz, I., et al. (2024). π-π Stacking constructing efficient charge channels for perovskite photovoltaics. Sci. Bull. *69*, 26-29, https://doi.org/10.1016/j.scib.2023.11.052.

35. Nematov, D., Kholmurodov, K., Husenzoda, M., Lyubchyk, A., and Burhonzoda, A. (2022). Molecular adsorption of H2O on TiO2 and TiO2: Y surfaces. Journal of Human, Earth, and Future, 3(2), 213–222. https://doi.org/10.28991/HEF-2022-03-02-07

36. Huang, L., Chen, Z., Wang, K. L., and Wang, Z. K. (2025). Dry-Processed Perovskite Photovoltaics: Materials, Fabrication Techniques, and Devices. Adv. Energy Mater. *e03501*, https://doi.org/10.1002/aenm.202503501.

37. Li, B., Li, S., Gong, J., Wu, X., Li, Z., Gao, D., Zhao, D., Zhang C., Wang. Y., and Zhu Z. (2024). Fundamental understanding of stability for halide perovskite photovoltaics: The importance of interfaces. Chem, *10*, 35-47, https://doi.org/10.1016/j.chempr.2023.09.002.

38. Tan, L., Zhou, J., Zhao, X., Wang, S., Li, M., Jiang, C., Li, H., Zhang, Y., Ye Y., Tress, W., et al. (2023). Combined vacuum evaporation and solution process for high-efficiency large-area perovskite solar cells with exceptional reproducibility. Adv. Mater., *35*, 2205027, https://doi.org/10.1002/adma.202205027.

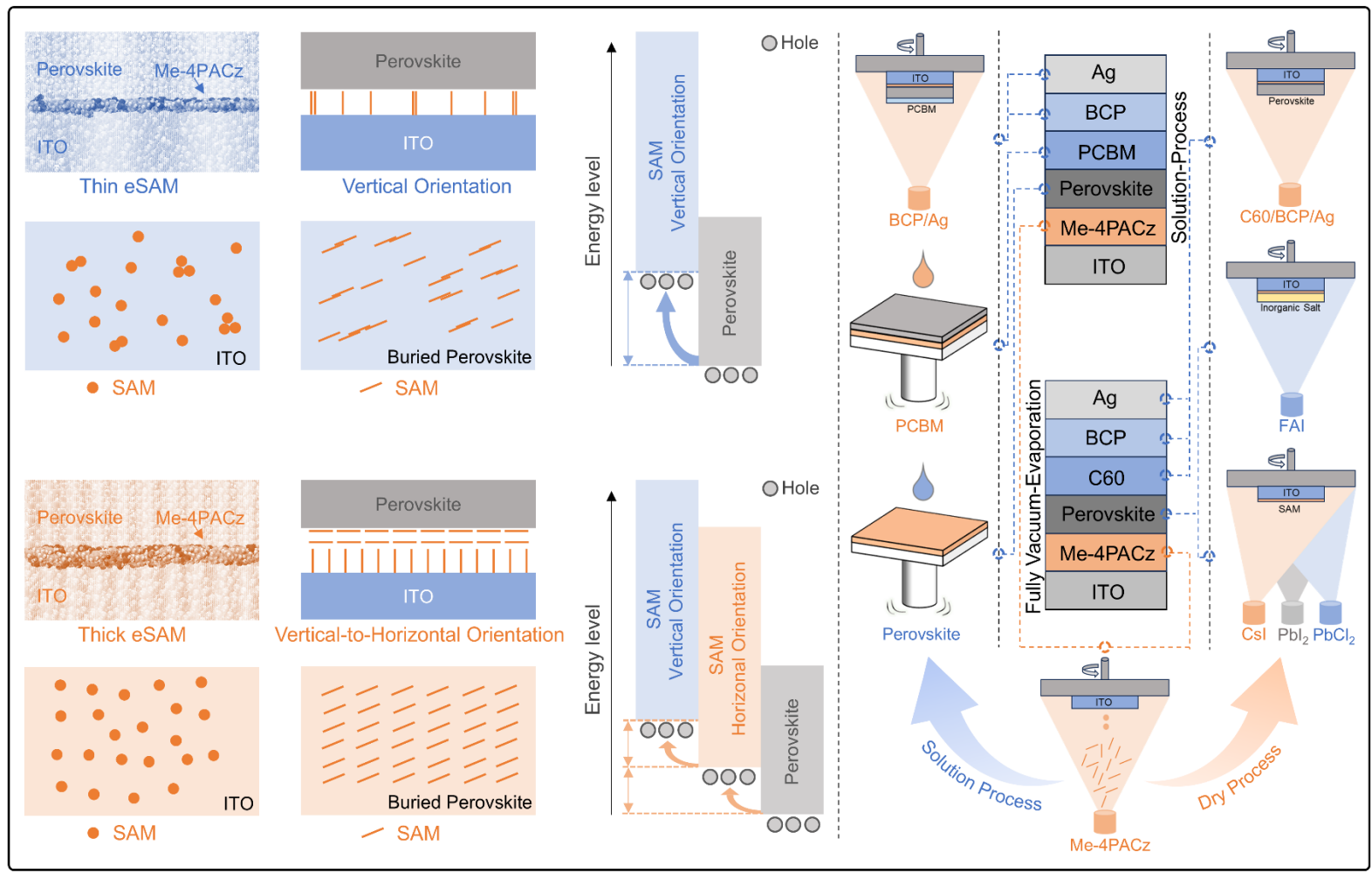

Figure 1. Schematic illustration of the thickness-regulation strategy for evaporated Me-4PACz and its application in perovskite solar cells.

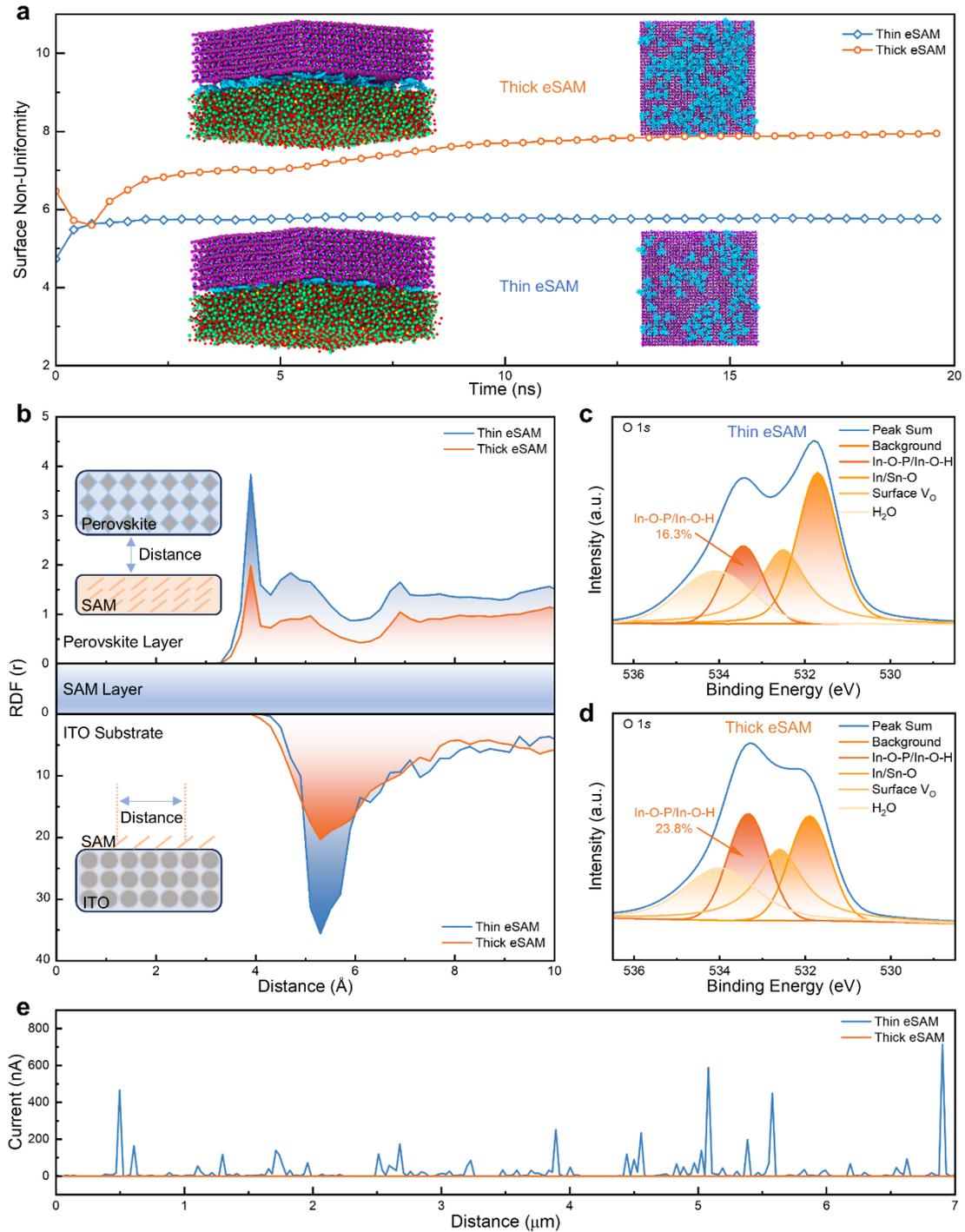

**Figure 2. Enhanced molecular coverage.**

(a) Molecular dynamics simulations for 20 ns to a very thin one layer of Me-4PACz (Thin eSAM) and a thicker two layer Me-4PACz (Thick eSAM).

(b) RDF analysis between SAM molecules and the perovskite surface (top) and RDF analysis between molecules in the SAM (bottom).

(c) and (d) XPS spectra of O 1$s$ for Thin eSAM and Thick eSAM.

(e) Current of the diagonal line scan in C-AFM images for Thin eSAM and Thick eSAM.

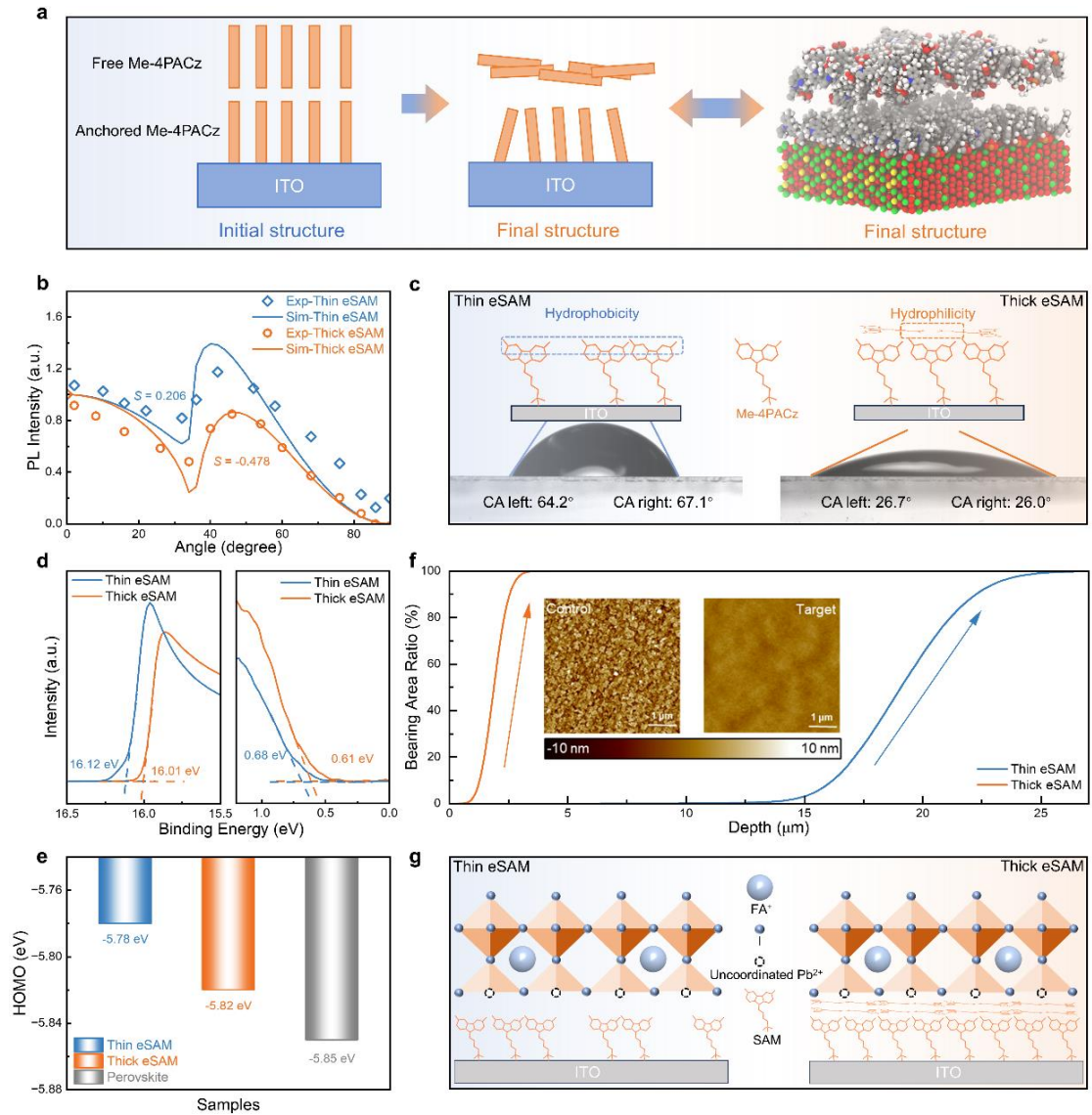

**Figure 3. Spontaneous vertical-to-horizontal orientation.**

(a) Molecular dynamics simulation for Me-4PACz orientation.

(b) Angle-dependent PL spectra for Thin eSAM and Thick eSAM.

(c) Water contact angle measurement for Thin eSAM and Thick eSAM.

(d) UPS spectra for Thin eSAM and Thick eSAM.

(e) Diagram of gradient energy level formation.

(f) AFM images of the Thin eSAM and Thick eSAM and corresponding bearing area ratio curves.

(g) Schematic illustration of Me-4PACz distribution between ITO and the perovskite layer.

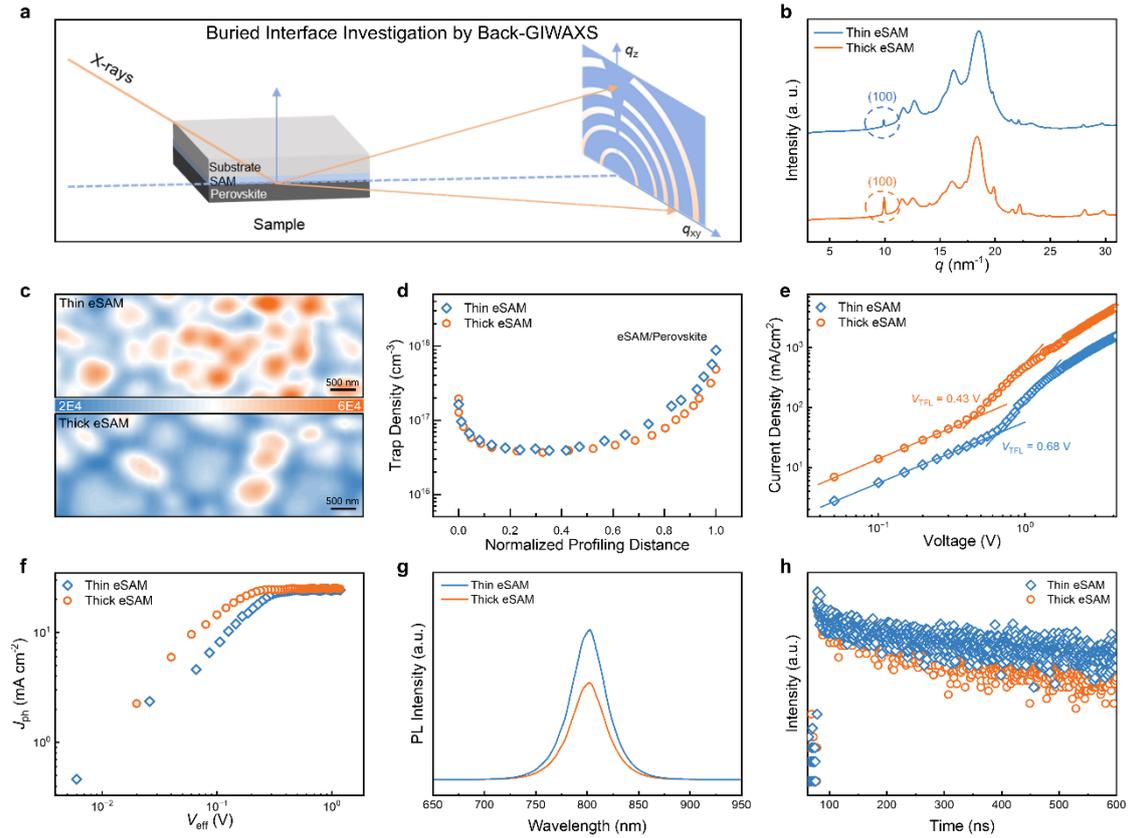

**Figure 4. Improved perovskite film quality.**

    (a) Schematic illustration for Back-GIWAXS technique.

    (b) Back-GIWAXS patterns of the perovskite films growing on Thin eSAM and Thick eSAM

    (c) PL mapping images for the perovskite films growing on Thin eSAM and Thick eSAM.

    (d) DLCP curves of devices based on Thin eSAM and Thick eSAM.

    (e) SCLC curves of hole-only devices based on Thin eSAM and Thick eSAM.

    (f) $J_{ph}$–$V_{eff}$ curves of devices based on Thin eSAM and Thick eSAM.

    (g) The PL spectra for perovskite films fabricated on Thin eSAM and Thick eSAM.

    (h) The TRPL decay curves for perovskite films fabricated on Thin eSAM and Thick eSAM.

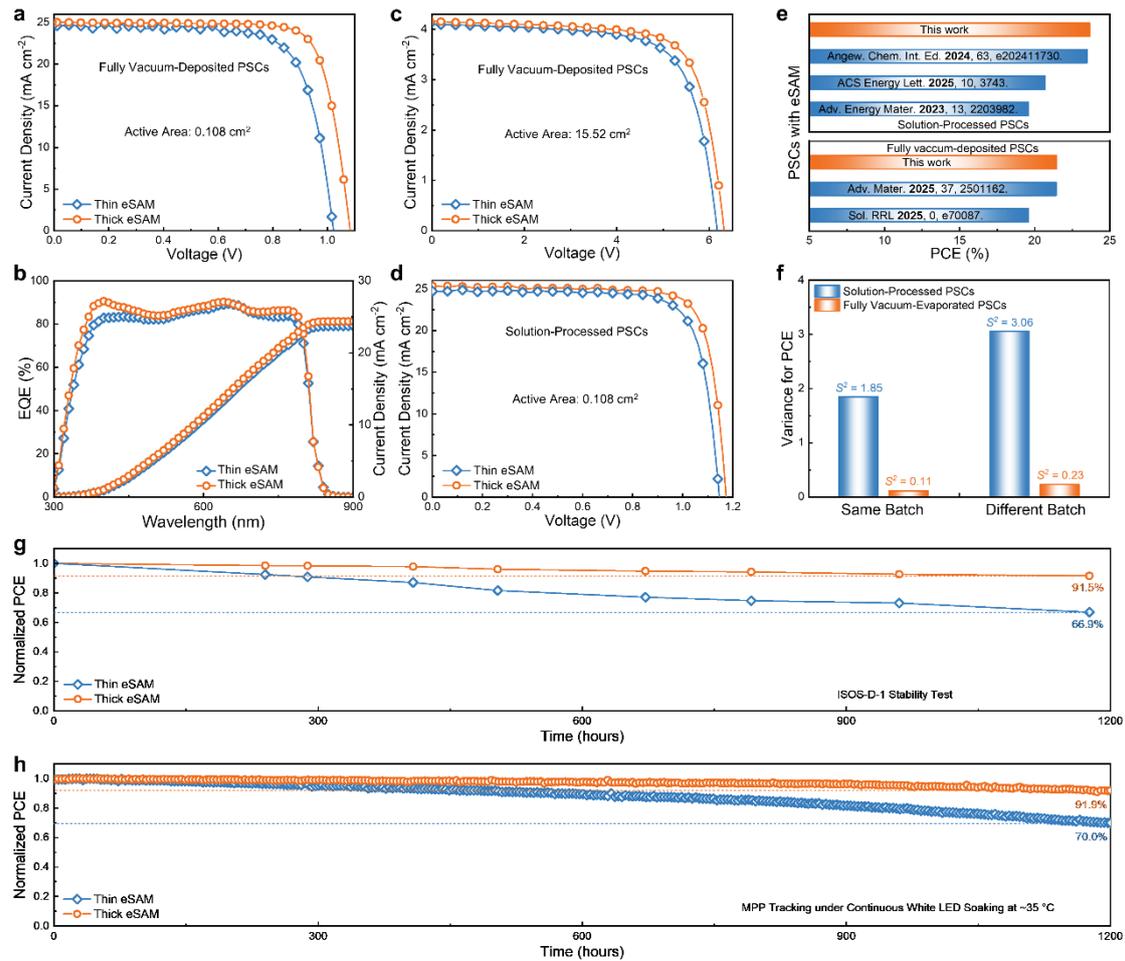

**Figure 5. Device performance and Stability.**

(a) *J-V* characteristic curves of small-area fully vacuum-deposited PSCs.

(b) EQE spectra of fully vacuum-deposited PSCs.

(c) *J-V* characteristic curves of large-area fully vacuum-deposited PSCs modules.

(d) *J-V* characteristic curves of solution-processed PSCs.

(e) Comparison of device performance reported in literature for fully vacuum-deposited PSCs and solution-processed PSCs fabricated using eSAM.

(f) Variance statistical plot of PCE for fully vacuum-deposited PSCs and solution-processed PSCs (Same Batch and Different Batch).

(g) PCE degradation curves of fully vacuum-deposited PSCs without encapsulation in an ambient condition for 1176 h.

(h) Continuous maximum power point tracking (MPPT) operational stability of fully vacuum-deposited PSCs (1-sun equivalent white LED illumination in $N_2$ atmosphere, ~35 °C).